\documentstyle[twocolumn,aps]{revtex}
\draft
\begin{document}

\title{Electronic and Magnetic Structure of LaCuO$_{2.5}$}

\author{B. Normand and T. M. Rice}

\address{Theoretische Physik, ETH-H\"onggerberg, CH-8093 Z\"urich, 
Switzerland.}

\date{\today}

\maketitle

\begin{abstract}

	The recently-discovered ``ladder'' compound LaCuO$_{2.5}$ has been
found to admit hole doping without altering its structure of coupled copper 
oxide ladders. While susceptibility measurements on the parent compound 
suggest a spin gap and a 
spin-liquid state, NMR results indicate magnetic order 
at low temperatures. These seemingly contradictory results may 
be reconciled if in fact the magnetic state is near the 
crossover from spin liquid to antiferromagnet, and we 
investigate this possibility. From a tight-binding fit to the valence 
LDA bandstructure, we deduce that the strength of the interladder hopping 
term is approximately half that of intraladder hopping, showing that the 
material is three-dimensional in character. A mean-field treatment 
of the insulating magnetic state gives a spin-liquid phase whose 
spin gap decreases with increasing interladder coupling, vanishing 
(signalling a transition to the ordered phase) at a value somewhat below 
that obtained for LaCuO$_{2.5}$. The introduction of an on-site 
repulsion term, $U$, to the band scheme causes a transition to an 
antiferromagnetic insulator for rather small but finite values of $U$, 
reflecting the predominance of (one-dimensional) ladder behavior, and 
an absence of any special nesting features. 

\end{abstract}

\pacs{PACS numbers: 75.10.Jm, 75.30.Kz, 75.40.Cx, 75.50.Ee }


\section{Introduction}

	One of the interesting and challenging sub-fields of low-dimensional 
quantum magnetism which has emerged from the wealth of activity directed at
improving the understanding of high-temperature superconductors is that of
ladder systems \cite{rdr}. These consist of $n$ parallel, interacting chains 
of $S$ = \mbox{$\frac{1}{2}$} ions, which can be considered as a spin ladder 
with $n$ legs, and rungs of $n$-1 bonds. The ladders have only weak mutual 
interactions. A combination of 
experimental and theoretical efforts has in the last few years produced 
significant advances in the realization and understanding of the properties 
of spin ladders, some of which are not at all intuitive.

	Ladder cuprates emerged first with the discovery by Hiroi {\it et
al.}\cite{rhatb} that in the series of materials Sr$_n$Cu$_{n+1}$O$_{2n+1}$ 
it is possible to create two-dimensional, stoichiometric copper oxide planes 
of composition Cu$_{n+1}$ O$_{2n+1}$ by removing from the uniform CuO$_2$ 
plane parallel, equally-spaced lines of oxygen atoms. It was pointed 
out by Rice {\it et al.}\cite{rrgs} that because these shear defects 
give rise to only weak, ferromagnetic interactions between neighboring 
copper spins, the remaining strips of CuO$_2$ plane will appear as isolated 
$n$+1-leg ladders of antiferromagnetically coupled spins. These authors
proposed that the systems should then illustrate the contrasting, and now
well-established properties of even- and odd-leg ladders, that the former show
a gap to spin excitations (spin gap) with consequent exponential spatial decay
of correlations, while the latter are gapless with power-law
decays. Subsequent susceptibility \cite{rahtik} and nuclear magnetic resonance
(NMR) \cite{rikaaht} experiments have amply borne out this conjecture.

	The theoretical understanding of ladders has been in large part 
based on numerical techniques, which are particularly well suited to systems
of such restricted dimensionality, and this is reviewed in Ref. \cite{rdr}. 
The first indications that the 2-leg ladder should exhibit a spin gap came 
from numerical work \cite{rdrs,rh}, and a variety of methods has since been 
applied to investigate the spin gap and spin correlations, and the properties 
of multi-leg
ladders.\cite{rbdrs,rwns,rfat} These studies not only confirm the picture of
the spin system emerging from analytic approaches \cite{rs} valid in certain 
limits, but provide the most accurate information available on the 
properties of this class of strongly-correlated systems.

	The second proposal concerning the properties of ladder compounds made
in Ref. \cite{rrgs}, that the doped ladder should become superconducting with 
a $d$-wave order parameter, has proved harder to test for materials reasons. 
However, the compound LaCuO$_{2.5}$, recently synthesized by Hiroi and 
Takano,\cite{rht} has been found to admit hole doping without altering its 
structure of coupled, 2-leg copper oxide ladders, and so constitutes the first 
case in which one may seek doping-dependent behavior analogous to that 
of the high-temperature superconductors. The high-pressure phase of 
La$_{1-x}$Sr$_x$CuO$_{2.5}$ is derived from a cubic, three-dimensional 
perovskite structure. The absence of oxygen atoms along lines 
leaves 2-leg ladders which relax from a relative angle of 90$^o$ in the 
primitive structure to 62$^o$ in the depleted one. The material is then 
orthorhombic (space group P$_{\rm bam}$), with four copper atoms per unit 
cell, and may be considered as a set of ladders 
oriented along the $z$-axis, and periodically arrayed in the $x$- and 
$y$-directions, as shown in Fig. 1(a). This structure is not altered with 
Sr doping to $x = 0.2$. The interladder coupling in the ($x,y$) 
plane arises because each oxygen atom at the outside edges of the 2-leg 
ladders, which is part of the planar CuO$_4$ unit, is also effectively 
apical to a copper atom in a neighboring ladder, contributing to a finite 
transfer integral. 

	Susceptibility measurements \cite{rht} on the parent compound up to
temperatures of 500K, interpreted by a formula proposed by Troyer {\it et 
al.},\cite{rttw} suggest the presence of a spin gap in the excitation 
spectrum, and therefore a spin liquid state. In contrast, NMR studies of 
the same samples \cite{rmkiahkt} indicate that the system orders 
antiferromagnetically at low temperatures, below an apparent $T_N \simeq 
117$K. Since the spin susceptibility should evolve continuously as one passes
through the quantum critical point separating the spin-liquid state and the
state with long-range antiferromagnetic order, it is possible that even in the
latter, close to the quantum critical point, the spin susceptibility will
decrease substantially as the temperature is lowered. Such behavior may be
difficult to distinguish from that of a true spin liquid with a finite gap. 
Therefore it is plausible that the seemingly contradictory experimental 
results may be reconciled if the magnetic state is close to the crossover from 
an antiferromagnet to the spin liquid. Here we seek evidence, by examining 
the effects of interladder interactions, that the system is indeed close to 
this quantum critical point. 

	The outline of this paper is as follows. In section II we
present a tight-binding fit to the LDA bandstructure to extract the
interladder hopping matrix elements, and use these to obtain the 
superexchange interactions. We consider in section III the nature of 
the spin-liquid ground state for ladders of spins coupled in both 
orthogonal directions, using a mean-field approach to estimate the location of
the quantum critical point. In section IV we introduce a double-occupancy term 
to the tight-binding bands, to investigate the degree of three-dimensional 
character in the electronic structure as a result of interladder interactions. 
Section V contains our conclusions and a brief discussion.

\section{Tight-Binding Fit to the LDA Bandstructure}

	The bandstructure of LaCuO$_{2.5}$ has been computed by 
Mattheiss,\cite{rm} using the Local Density Approximation (LDA) method. Here 
we examine the dispersion of only the highest occupied valence bands,
and use a tight-binding model based on the single, copper-based orbital in 
each planar CuO$_4$ unit within the ladder which lies closest to the chemical 
potential: in the Cu$_{n+1}$O$_{2n+1}$ strip this is the antibonding, 
Cu-centered 3$d_{x^2 - y^2}$ orbital. Restricting the set of hopping matrix 
elements to those between nearby copper atoms, we obtain an effective
one-band model, albeit with four mixed levels (one from each Cu atom in the 
unit cell). These generate a complex of four bands, which lie close to or cut 
the Fermi energy. The complex is 
half-filled in the undoped system, and well separated from other bands on 
the small energy scales of most physical interest, so can be taken to 
determine the low-energy behavior of the model. With this interpretation, 
one may deduce the ratio $t^{\prime} / t$, of inter- and intraladder atomic 
orbital overlap, and thus estimate the ratio $J^{\prime} / 
J$ of the magnetic interactions, by using the superexchange result
\cite{ra} $J \simeq 4 t^2 / U$.

	The tight-binding Hamiltonian is 
\begin{equation}
H = - \sum_{ij \sigma} t_{ij} c_{i \sigma}^{\dag} c_{j \sigma} ,
\label{esh}
\end{equation}
in which $i$ and $j$ each denote a pair ($n,m$), where $n$ labels the unit
cell, and $m = 1, \dots 4$ the different atoms within each cell. The hopping 
matrix elements $t_{ij}$ may be taken to be short-ranged, and the maximal set 
which we need to achieve a reasonable fit is illustrated in Fig. 1(b). 
Starting from tight-binding parameter fits in the two-dimensional CuO$_2$
plane,\cite{rfg,rnkf} we choose the nearest-neighbor intraladder parameters, 
$t_r$ for hopping along a rung and $t_z$ for hopping along a leg, to have the 
same value $t_r = t_z = t \simeq 0.4$eV. We use as the next-nearest-neighbor
parameters $t_{r}^{\prime} = - $\mbox{$\frac{1}{5}$}$ t_r$ and $t_{z}^{\prime} 
= \, $\mbox{$\frac{1}{6}$}$ t_z$ for hopping across a square plaquette in the 
ladder, and along two leg bonds, respectively. Finally, we introduce a 
parameter $t_s$ for hopping from a copper atom on one ladder to its neighbor 
on the adjacent ladder, and also the next-neighbor analog
$t_{s}^{\prime}$ for transfer between atoms on adjacent ladders with a
relative displacement of one leg bond. Because of the low symmetry of the 
LaCuO$_{2.5}$ structure, the Cu-(apical) O-Cu interladder bond is far from 
straight, with the CuO$_5$ pyramids quite irregular, and this distortion 
may allow a substantial value of $t_s$. The role of each of the terms in 
the fitting scheme will be illustrated below. The next-neighbor parameters
$t_{r,s,z}^{\prime}$ are expected to be significantly smaller than
their direct counterparts, due to the short-range nature of the
overlaps, and the important parameter to fit will be the ratio of $t_s$ 
to $t_r$.

	The Hamiltonian (\ref{esh}) may be expressed in matrix form in
reciprocal space as 
\begin{equation}
H = \sum_{{\bf k} \sigma} {\bf c}_{{\bf k} \sigma}^{\dag} {\bf H}_{{\bf k}} 
{\bf c}_{{\bf k} \sigma}, 
\label{esmh}
\end{equation}
where ${\bf c}_{{\bf k} \sigma}^{\dag} = (c_{{\bf k} \sigma}^{1 \dag}, 
c_{{\bf k} \sigma}^{2 \dag}, c_{{\bf k} \sigma}^{3 \dag}, c_{{\bf k} 
\sigma}^{4 \dag})$ is the 
vector of creation operators 
\begin{equation}
c_{{\bf k} \sigma}^{m \dag} = \frac{1}{\sqrt{N}} \sum_n 
{\rm e}^{i {\bf k.R}_{n,m}} 
c_{n,m \sigma}^{\dag}
\label{ebsco}
\end{equation}
for the Bloch states formed by separate linear combination of the 
$d_{x^2 - y^2}$ orbitals of each copper atom in the unit cell, 
and ${\bf k}$ is a vector in the orthorhombic Brillouin zone. Setting the
lattice constants $a$, $b$ and $c$ to unity, the Hamiltonian matrix is 
\begin{equation}
{\bf H}_{\bf k} = - \left( \begin{array}{cccc}
t_z (k_z) & {\overline t}_s {\rm e}^{i k_y s_y} & 0 & {\tilde t}_r {\rm e}^{-
i {\bf k}.{\bf {\overline r}}} \\ {\overline t}_s {\rm e}^{- i k_y s_y} & 
t_z (k_z) & {\tilde t}_r {\rm e}^{i {\bf k}.{\bf r}} & 0 \\ 0 & {\tilde t}_r 
{\rm e}^{- i {\bf k}.{\bf r}} & t_z (k_z) & {\overline t}_s {\rm e}^{i k_y 
s_y} \\ {\tilde t}_r {\rm e}^{i {\bf k}.{\bf {\overline r}}} & 0 & 
{\overline t}_s {\rm e}^{- i k_y s_y} & t_z (k_z) \end{array} \right) ,
\label{eshm}
\end{equation}
where $t_z (k_z) = 2 t_z \cos k_z + 2 t_{z}^{\prime} \cos 2 k_z$, 
${\overline t}_s = 2 {\tilde t}_s \cos $\mbox{$\frac{1}{2}$}$ k_x$, in
which the factor $\cos $\mbox{$\frac{1}{2}$}$ k_x$ arises because $s_x = 0.5$, 
${\tilde t}_{\nu} = t_{\nu} + 2 t_{\nu}^{\prime} \cos k_z$, and the bond 
vectors ${\bf r}$(${\bf {\overline r}}$) and ${\bf s}$(${\bf {\overline s}}$) 
are shown in Fig. 1(c). The eigenvalue problem
gives an equation quadratic in the squares of the mode frequencies, whose
solutions are the dispersion relations of the four energy bands
\begin{eqnarray}
\epsilon_{\bf k} & = & \pm \left[ {\tilde t}_{r}^{2} + 4 {\tilde t}_{s}^{2} 
\cos^2 \mbox{$\frac{1}{2}$} k_x \pm 4 {\tilde t}_{r} 
{\tilde t}_{s} \cos \mbox{$\frac{1}{2}$} k_x \cos \mbox{$\frac{1}{2}$} 
k_y \right]^{1/2} \label{eqpb} \nonumber \\ & & \;\;\;\; - 2 t_z 
\cos k_z - 2 t_{z}^{\prime} \cos 2 k_z .
\end{eqnarray}
This concise, closed form emerges because the exponential factors ${\rm
e}^{\pm (2 i k_y r_y + 2 i k_y s_y)}$ may be collected as $1 - \cos k_y$, 
as $r_y + s_y = 0.5$. The general dispersion simplifies further for
particular values of $k_x$ and $k_y$ in the Brillouin zone: in particular, on
the zone faces ($k_x, k_y = \pi$), the last term in the square root vanishes
and the bands are doubly degenerate, as required by the group-theoretical
analysis of structures with non-symmorphic space groups. 

	The dispersions of the 4 bands are shown in Fig. 2(a) for a series of
high-symmetry lines in the orthorhombic Brillouin zone, and for the ratio $t_s
/ t_r = 0.4$. The labelling of
points is shown in Fig. 2(c), and their order is chosen to match the results
of Mattheiss.\cite{rm} Also shown in the figure are the energy-spacings
dictated by the choice of the tight-binding parameters, as these vary between
their maximum and minimum values along the chosen directions. 

	For an isolated ladder along ${\bf {\hat z}}$, i.e. with no 
interladder interactions (${\tilde t}_s = 0$), the bands would be completely 
flat around $\Gamma$XSY$\Gamma$ and ZURTZ, with only a cosine dispersion 
(mildly perturbed by the $t_{z}^{\prime}$ term) along $\Gamma$Z. In this
situation there would be two doubly-degenerate bands around the zone center
and zone face, corresponding to the bonding and antibonding bands of each
ladder. These features are reflected in the two branches observed in Fig. 
2(a). From Eq. (\ref{eqpb}), the separation of the degenerate bands along the 
zone face XSY and edge URT is governed by the combinations $t_r
\pm 2 t_{r}^{\prime}$, and the splitting of the degeneracy along $\Gamma$X,
$\Gamma$Y and ZU, ZT by the combinations $t_s \pm 2
t_{s}^{\prime}$. Considering first the intraladder parameters $t_z$,
$t_{z}^{\prime}$, $t_r$ and $t_{r}^{\prime}$ chosen from the two-dimensional 
CuO$_2$ plane, we find good agreement of the tight-binding result
with that from LDA. Particularly notable is that the negative sign of 
$t_{r}^{\prime}$, and the relative magnitude \mbox{$\frac{1}{5}$}$ t_r$, are
required to reproduce the bands separations along XSY and URT. There 
is no evidence that a value of $t_r$ different from $t_z$ would improve the 
fit. The next-neighbor hopping parameter along the ladder, $t_{z}^{\prime}$,
appears only as an asymmetry of the cosine dispersion along $\Gamma$Z, and
the value chosen is in qualitative accord with the LDA result. That 
CuO$_2$ plane parameters remain appropriate for the ladder confirms the 
predominantly local picture of the interactions between copper sites. Turning 
to the band splitting at the $\Gamma$ and Z points, we find that the relatively
large value $t_s = \, $\mbox{$\frac{1}{2}$}$ t_r$ (Fig. 2(b)) gives the best 
qualitative reproduction of the bands crossing the chemical potential in the 
full LDA calculation by Mattheiss,\cite{rm} but that the value $t_s = \,
$\mbox{$\frac{2}{5}$}$ t_r$ appears closer to the results of a ``12-parameter 
fit'' illustrated in the same reference. The difference of the $\Gamma$ and
Z splittings is given rather well by the starting choice of $t_{s}^{\prime}
= \, $\mbox{$\frac{1}{5}$}$ t_s$. 

	It is clear from Fig. 2 that the primary feature of the dispersion
remains that in the $k_z$ direction, i.e. along the ladder. The Fermi 
surfaces for each band are determined almost exclusively by this part of the 
dispersion, in that they appear as sheets which are almost flat, perturbed 
little by the $t_r$ and $t_s$ terms, and have $k_z$ as normal. At half-filling,
the lowest band is an exception to this situation, because part of it is also 
filled  close to the Z point. On doping with holes, this region is rapidly 
emptied (below 5\%) so that all four Fermi surfaces are sheets parallel to the 
($k_x,k_y$) plane. Only when the doping level reaches 20\% does the chemical 
potential drop below the highest band in the $\Gamma$XSY$\Gamma$ plane, 
causing a pocket to open at the $\Gamma$ point for this band. 

	In summary, we find that a simple model of a single orbital per copper 
atom provides a good fit to the bandstructure. While there is scope for some 
variation in the choice of intraladder parameters, with this level of 
agreement between the tight-binding results and those of LDA it is not 
worthwhile to optimize further. 
We choose to work with the above values of the intraladder hopping matrix 
elements, and with the interladder overlap $t_s = \, $\mbox{$\frac{1}{2}$}$ 
t_r$, bearing in mind that this latter will be close to the upper limit of the 
narrow range of probable values. Estimating the superexchange interaction by 
$J^{\prime} \propto 4 t^{\prime 2} / U$ leads us to conclude that the 
interaction between spins on neighboring ladders will have a
magnitude $J^{\prime} \simeq 0.25 J$, where $J$ is the intraladder magnetic
coupling of both rung and leg spins. Because $J^{\prime}$ is an 
appreciable fraction of $J$, it is clear that the spin interactions in
LaCuO$_{2.5}$ will have significant three-dimensional character. 

\section{Mean-Field Analysis of the Spin Ground State}

	 A mean-field analysis of the spin state for ladder systems was
introduced by Gopalan {\it et al.} \cite{rgrs}, and in this section we follow 
closely the treatment of these authors. They employed a bond operator
representation of $S = \, $\mbox{$\frac{1}{2}$} quantum spins, used initially 
\cite{rsb} to investigate dimerized spin phases in two-dimensional 
systems, exploiting the fact that the topology of the ladder favors 
dimerization, particularly when the spin interaction, $J$, on a rung 
exceeds that on a leg, $\lambda J$. The authors proceeded in the mean-field
approximation to consider first the properties of an isolated ladder as a
function of the interaction ratio $\lambda$, then of periodic arrays of 
ladders in two dimensions, and finally of an array of frustrated double 
ladders of the type found in the SrCu$_2$O$_3$ system.\cite{rhatb} This 
approach can be considered to be exact in the limit where $\lambda 
\rightarrow 0$ and the spins form dimers on every rung, while its accuracy 
will diminish on extrapolating through finite $\lambda$ towards the desired 
isotropic point, $\lambda = 1$. 

	We begin by representing the system of spin ladders, each coupled by
two identical bonds per spin to separate, neighboring ladders, in a
geometrically simpler but topologically identical form, as shown in Fig. 3. 
Note that the spin configuration for local, antiferromagnetic interactions 
is unfrustrated, so a transition to an ordered antiferromagnet may be 
expected with increasing interladder coupling. 
The magnetic interactions are taken to be $J$ for spins on the same rung, 
$\lambda J$ for spins separated by a leg bond, and $\lambda^{\prime} J$ for
neighboring spins on different ladders, while no other couplings are
considered. The Hamiltonian for the spins is 
\begin{eqnarray}
H & = & J \sum_{j} {\bf S}_{l,j} {\bf .S}_{r,j} + \lambda J \sum_{j, m = l,r} 
{\bf S}_{m,j} {\bf .S}_{m,j+{\hat z}} \label{essh} \nonumber \\ 
& & + \lambda^{\prime} J \sum_{j} \left(
{\bf S}_{r,j} {\bf .S}_{l,j + \mbox{$\frac{1}{2}$} {\hat x} + 
\mbox{$\frac{1}{2}$} {\hat y}} + {\bf S}_{r,j} {\bf .S}_{l,j + 
\mbox{$\frac{1}{2}$} {\hat x} - \mbox{$\frac{1}{2}$} {\hat y}} \right) , 
\end{eqnarray}
where $j$ is a rung bond index and 
the labels $l$ and $r$ denote spins on the left and right sides of the
ladder. Following Ref. \cite{rgrs}, transformation to the bond-operator
representation yields 
\begin{equation}
H = H_0 + H_1 + H_2 + H_{ho},
\label{ebosh}
\end{equation}
where
\begin{eqnarray}
H_0 & = & J \sum_{j,\alpha} \left( - \mbox{$\frac{3}{4}$} s_{j}^{\dag} s_{j} + 
\mbox{$\frac{1}{4}$} t_{j,\alpha}^{\dag} t_{j,\alpha} \right) \label{ebosh0}
\nonumber \\ & & - \sum_{j,\alpha} \mu_j \left( s_{j}^{\dag} s_{j} + 
t_{j,\alpha}^{\dag} t_{j,\alpha} - 1 \right) ,
\end{eqnarray}
\begin{equation}
H_1 = \mbox{$\frac{1}{2}$} \lambda J \sum_{j,\alpha} \left( t_{j,
\alpha}^{\dag} t_{j+{\hat z},\alpha} s_{j+{\hat z}}^{\dag} s_{j} + 
t_{j,\alpha}^{\dag} t_{j+{\hat z},\alpha}^{\dag} s_{j} s_{j+{\hat z}} 
+ H.c. \right) 
\label{ebosh1}
\end{equation}
and
\begin{eqnarray}
H_2 & = & - \mbox{$\frac{1}{4}$} \lambda^{\prime} J \sum_{j,\alpha} 
\sum_{\eta = \pm 1} \left( t_{j,\alpha}^{\dag} t_{j + \mbox{$\frac{1}{2}$} 
{\hat x} + \eta \mbox{$\frac{1}{2}$} {\hat y},\alpha} s_{j + 
\mbox{$\frac{1}{2}$} {\hat x} + \eta \mbox{$\frac{1}{2}$} {\hat y}}^{\dag} 
s_j \right. \label{ebosh2} \nonumber \\ & & \;\;\;\;\;\; + \left. 
t_{j,\alpha}^{\dag} t_{j + \mbox{$\frac{1}{2}$} {\hat x} + \eta 
\mbox{$\frac{1}{2}$} {\hat y},\alpha}^{\dag} s_{j + \mbox{$\frac{1}{2}$} 
{\hat x} + \eta \mbox{$\frac{1}{2}$} {\hat y}} s_j + H.c. \right) . 
\end{eqnarray}
In these equations, $s_{j}^{\dag}$ is the creation operator for a spin singlet
on bond $j$, the operators $t_{j,\alpha}^{\dag}$ create the three possible
triplet states on the same bond, and the Lagrange multiplier, $\mu_j$, 
introduced to ensure the constraint 
\begin{equation}
s_{j}^{\dag} s_j + \sum_{\alpha} t_{j,\alpha}^{\dag} t_{j,\alpha} = 1  
\label{ec}
\end{equation}
on each bond, which restricts the physical spin states to singlets or
triplets, appears as an effective chemical potential. 
The part $H_{ho}$ in (\ref{ebosh}) contains terms with three and
four $t_{j,\alpha}$ operators, and will be neglected in our approximation; 
in Ref. \cite{rgrs} it was shown in addition that the effects of
such higher-order terms are small. 

	Because the singlet on each bond has the lowest energy, we assume that
the system condenses into this state, leading to a finite expectation
value of the bosonic $s_j$ operator, $\langle s_j \rangle = {\overline s}$. 
This is the average expectation value, or mean-field value of the
operators $s_j$, and the site-dependent chemical potential $\mu_j$ is also
replaced by a global average value $\mu$. Working with the physical unit cell
(Fig. 3), 
which contains two rungs, we may transform the operators $t_{j,\alpha}$ in 
$H_0$, $H_1$ and $H_2$ to those for two types of triplets, $t_{k,\alpha}^1$ 
and $t_{k,\alpha}^2$. The Hamiltonian in this approximation is \cite{mzhc}
\begin{eqnarray}
H_{\rm m} (\mu, {\overline s}) & = & N \left( - \mbox{$\frac{3}{4}$} J
{\overline s}^2 - \mu  {\overline s}^2 + \mu \right) + \sum_{{\bf k} 
\alpha} \left\{ \sum_{\nu = 1,2} \left[ \Lambda_{\bf k} t_{{\bf k} 
\alpha}^{\nu \dag} t_{{\bf k} \alpha}^{\nu} \right. \right. \label{emfsh} 
\nonumber \\ & & \left. + \Delta_{\bf k} \left( t_{{\bf k} 
\alpha}^{\nu \dag} t_{-{\bf k} \alpha}^{\nu \dag} + t_{{\bf k} \alpha}^{\nu} 
t_{-{\bf k} \alpha}^{\nu} \right) \right]  + \left[ \Lambda_{\bf k}^{\prime} 
t_{{\bf k} \alpha}^{1 \dag} t_{{\bf k} \alpha}^{2} \right. \\ & & 
\left. \left. + \Delta_{\bf k}^{\prime} \left( t_{{\bf k} \alpha}^{1 \dag} 
t_{-{\bf k} \alpha}^{2 \dag} + t_{{\bf k} \alpha}^{1} t_{-{\bf k} 
\alpha}^{2} \right) \right] + [ 1 \leftrightarrow 2] \right\} , \nonumber
\end{eqnarray}
in which 
\begin{equation} 
\Lambda_{\bf k} = \mbox{$\frac{1}{4}$} J - \mu + J {\overline s}^2 \lambda 
\cos k_z ,
\label{elk}
\end{equation}
\begin{equation} 
\Delta_{\bf k} = \mbox{$\frac{1}{2}$} J {\overline s}^2 \lambda \cos k_z ,
\label{edk}
\end{equation}
\begin{equation} 
\Lambda_{\bf k}^{\prime} = 2 \Delta_{\bf k}^{\prime} =  - 
J^{\prime} {\overline s}^2  \cos \mbox{$\frac{1}{2}$} k_x \cos 
\mbox{$\frac{1}{2}$} k_y ,
\label{elddk}
\end{equation}
and $N$ denotes the total number of ladder rungs. 
The part of $H_{\rm m}$ (\ref{emfsh}) dependent on the triplet operators
is diagonalized by the non-unitary, bosonic Bogoliubov transformation 
\begin{equation} 
\gamma_{{\bf k} \alpha}^{\pm} = \cosh \theta_{\bf k}^{\pm} \left( t_{{\bf k} 
\alpha}^{1} \pm t_{{\bf k} \alpha}^{2} \right) + \sinh \theta_{\bf k}^{\pm} 
\left( \pm \, t_{-{\bf k} \alpha}^{1 \dag} - t_{-{\bf k} \alpha}^{2 \dag}
\right), 
\label{ebt}
\end{equation}
whose coefficients are given by  
\begin{equation}
\cosh 2 \theta_{\bf k}^{\pm} = \frac{\Lambda_{\bf k} \pm \Lambda_{\bf
k}^{\prime}}{\omega_{\bf k}^{\pm}} \;\;\;\; \sinh 2 \theta_{\bf k}^{\pm} 
= \frac{2 \left( \Delta_{\bf k} \pm \Delta_{\bf k}^{\prime} \right)}
{\omega_{\bf k}^{\pm}} ,
\label{ebtc} 
\end{equation}
where in turn 
\begin{equation} 
\omega_{\bf k}^{\pm} = \sqrt{ \left( \Lambda_{\bf k} \pm \Lambda_{\bf
k}^{\prime} \right)^2 - 4 \left( \Delta_{\bf k} \pm \Delta_{\bf k}^{\prime}
\right)^2 } 
\label{eqd}
\end{equation}
yield the dispersion relations of the two magnon branches. The Hamiltonian 
now takes the form 
\begin{eqnarray}
H_{\rm m} (\mu, {\overline s}) & = & N \left( - \mbox{$\frac{3}{4}$} J
{\overline s}^2 - \mu  {\overline s}^2 + \mu \right) - \mbox{$\frac{3}{2}$} 
N \left( \mbox{$\frac{1}{4}$} J - \mu \right) \label{edmfsh} \nonumber \\ & & 
+ \sum_{{\bf k} \alpha} \sum_{\nu = \pm} \omega_{\bf k}^{\nu} \left( 
\gamma_{{\bf k} \alpha}^{\nu \dag} \gamma_{{\bf k} \alpha}^{\nu} + 
\mbox{$\frac{1}{2}$} \right) , 
\end{eqnarray}
which contains the mean-field part and the zero-point quantum corrections 
from the triplet magnon excitations. The mean-field equations to be solved 
self-consistently for $\mu$ and ${\overline s}$ are given by 
\begin{equation} 
\langle \frac{\partial H_{\rm m}}{\partial \mu} \rangle = 0 = {\overline s}^2 
- \frac{5}{2} + 3 \sum_{{\bf k} \; \nu = \pm} \frac{\left( \Lambda_{\bf 
k} + \nu \Lambda_{\bf k}^{\prime} \right) }{4 \omega_{\bf k}^{\nu}} n_m 
(\omega_{\bf k}^{\nu})
\label{emfem}
\end{equation}
and 
\begin{equation} 
\langle \frac{\partial H_{\rm m}}{\partial {\overline s}} \rangle = 0 = 
\frac{3}{2} + 2 \frac{\mu}{J} - 3 \sum_{{\bf k} \; \nu = \pm} 
\frac{\Lambda_{\bf k} - 2 \Delta_{\bf k}}{2 \omega_{\bf k}^{\nu}} 
a_{\bf k}^{\nu} n_m (\omega_{\bf k}^{\nu}) ,
\label{emfes}
\end{equation}
where
\begin{equation}
a_{\bf k}^{\pm} = \lambda \cos k_z \pm \lambda^{\prime} \cos 
\mbox{$\frac{1}{2}$} k_x \cos \mbox{$\frac{1}{2}$} k_y 
\label{eak}
\end{equation}
contain the dispersive parts of $\omega_{\bf k}^{\pm}$. $n_m (\omega_{\bf
k}^{\pm})$ denotes the magnon thermal occupation function, and will be 
discussed in more detail in a future publication. 
The factor of 3 preceding the ${\bf k}$-summations in both equations is the 
result of the sum over $\alpha$ for the three triplet magnon 
states.\cite{rzupc} This factor was omitted in Ref. \cite{rgrs}, and we 
comment below on the effect of the correction on the results presented there. 

	The mean-field equations are solved at zero temperature, where the
thermal factor becomes unity, and by taking the continuum limit in which the
${\bf k}$-sum becomes an integral over three-dimensional reciprocal space. 
As in Ref. \cite{rgrs} we reduce the two equations to a single one for the 
variable 
\begin{equation}
d = \frac{2 J {\overline s}^2}{\mbox{$\frac{1}{4}$} J - \mu} , 
\label{ed}
\end{equation}
which has the form 
\begin{equation}
d = 5 - \mbox{$\frac{3}{2}$} \sum_{\nu = \pm} \int \frac{d^3 k}{(2 \pi)^3} 
\frac{1}{\sqrt{1 + d a_{\bf k}^{\nu}}} . 
\label{escd}
\end{equation}
As a characteristic 
parameter of the spin-liquid ground state, we will be most interested in 
the value of the spin gap, the minimum excitation energy of the lower branch
of the triplet magnon excitations, which is given by 
\begin{equation}
\Delta = \left( \mbox{$\frac{1}{4}$} J - \mu \right) \sqrt{1 - d \left( 
\lambda + \lambda^{\prime} \right)} , 
\label{esg}
\end{equation}
with $d$ determined by the mean-field equation
(\ref{escd}). From Eq. (\ref{eqd}) we see that the excitation spectrum has a
minimum at the wavevector ${\bf k}_M = (0,0,\pi)$ in the reciprocal lattice 
of the bipartite structure shown in Fig. 3. 
The value of $\lambda^{\prime}$ where $\Delta$ is driven to zero will
give the transition from the spin liquid state, where the spin orientation
fluctuates with a time-averaged value of zero and with short-range 
correlations primarily at ${\bf k}_M$, to a magnetically ordered 
state characterized by ${\bf k}_M$. This wavevector corresponds to uniform 
polarization of the spin singlets on ladder rungs in the ($x,y$) 
plane, with spins oppositely directed between neighboring planes in the 
$z$-direction, i.e. a simple antiferromagnetically aligned spin
pattern. 

	In the limit of no interladder coupling we obtain the spin gap 
$\Delta_0 = 0.501 J$ of the isolated, isotropic ($\lambda = 1$) 2-leg ladder. 
This is only a mean-field result, but is in very good agreement with the 
result $\Delta_0 = 0.504 J$ of numerical studies \cite{rwns} by the Density 
Matrix Renormalization Group technique. In fact this agreement 
is largely serendipitous, and deteriorates on taking into account the 
higher-order terms \cite{rzupc}; in the mean-field approximation, the spin gap
of the isolated ladder diverges logarithmically in the limit of large
$\lambda$,\cite{rgrs} and the effects of this increase are already manifest at
$\lambda = 1$, causing the mean-field result, which initially underestimates
the spin gap, to recover towards the exact value determined numerically. In 
this treatment we have
not had to invoke a self-energy correction term: in Ref. \cite {rgrs}, the
authors investigated the curious qualitative behavior of the solution to their 
(erroneous) mean-field equation by expanding in small $\lambda$ about the limit
of strong rung coupling where the dimer treatment is accurate. In the
corrected mean-field theory, one obtains 
\begin{equation}
\Delta = J \left( 1 - \lambda + \mbox{$\frac{1}{4}$} \lambda^2 + 
\mbox{$\frac{3}{8}$} \lambda^3 + O \left( \lambda^4 \right) \right) ,  
\label{esge}
\end{equation}
which corresponds reasonably well to the result 
\begin{equation}
\Delta = J \left( 1 - \lambda + \mbox{$\frac{1}{2}$} \lambda^2 + 
\mbox{$\frac{1}{4}$} \lambda^3 + O \left( \lambda^4 \right) \right)   
\label{esgrtr}
\end{equation}
of a detailed strong-coupling analysis including excitation modes.\cite{rrtr}
Previously, the coefficient of the quadratic term had been found to be
negative, and so a self-energy term $\beta \lambda^2$ was introduced to 
correct for short-range interaction effects which appeared to have been
missing at the mean-field level; the chosen value $\beta = 0.7$ brought the
results into good argeement with previous numerical ones, and with the above
approximate treatments. 

	The spin gap obtained from the solution of (\ref{escd}) for the
three-dimensionally coupled ladder system is shown in Fig. 4(a) as a function 
of the interladder coupling $\lambda^{\prime}$, for fixed $\lambda = 1$. We 
see immediately that the spin gap decreases monotonically, with the 
transition point at $\lambda^{\prime} = 0.121$. Comparison with the result of 
section II indicates that the LaCuO$_{2.5}$ system should lie within the
ordered antiferromagnetic regime, but that it is indeed located in the
vicinity of the quantum critical point marking the phase transition from 
spin liquid to magnetic order. 

	It is also instructive to compare the appearance of the spin gap 
with that in a two-dimensional, unfrustrated, periodic array of coupled 
ladders.\cite{rgrs} In this case there is only one magnon branch, and the 
dispersive factor in the excitation spectrum is  
\begin{equation}
a_{\bf k} =  \lambda \cos k_z - \mbox{$\frac{1}{2}$} \lambda^{\prime} 
\cos k_x ,
\label{eak2}
\end{equation}
where $k_x$ is a wavevector parallel to a ladder rung. Again ${\bf k}_M = 
(0,\pi)$ corresponds to a simple antiferromagnetic spin alignment in the 
ordered phase, and the factor of \mbox{$\frac{1}{2}$} appears because there 
is only one bond per spin between ladders in such a system. The spin gap for 
the two-dimensional array is shown as a function of $\lambda^{\prime}$ in
Fig. 4(b), where again $\lambda = 1$ and the gap at $\lambda^{\prime} = 0$ 
is that of the isolated, isotropic ladder. The critical value of the 
interladder coupling, $\lambda_{c}^{\prime} = 0.43$, is seen to be
significantly greater than twice that in the three-dimensional case above, 
as might be expected for the simple reason that there are half as many 
interladder interactions, which may be taken as an indication that the spin
liquid state is more robust in lower dimensions. The almost linear decrease, 
in contrast to the downward curvature of the function in Fig. 4(a), 
illustrates a further effect of dimensionality, and agrees well with
the qualitative result of Ref. \cite{rgrs}.

	We may conclude that to within the accuracy of the mean-field 
approach for an isotropic ladder system, the above analysis provides good 
evidence that LaCuO$_{2.5}$ lies on the magnetically ordered side of the 
quantum critical regime of the transition between spin liquid and
antiferromagnet. 

\section{Hartree-Fock Approximation to the Hubbard Model}

	We have seen in the previous sections that the interladder 
interactions in LaCuO$_{2.5}$ are quite strong, and possibly sufficiently 
strong to change the magnetic structure from a spin liquid to a 
three-dimensionally ordered antiferromagnet. To examine further the 
character of the electronic structure, we introduce 
an on-site Coulomb interaction, $U$, to the tight-binding bandstructure 
of section II, and perform a Hartree-Fock calculation of an ordered 
antiferromagnetic state. In a strictly one-dimensional system, the critical 
value of $U$ necessary to stabilize an insulating antiferromagnet vanishes 
at half-filling, whereas in a general, three-dimensional electronic structure 
it is of the order of the bandwidth. 

	Introduction of the on-site Coulomb interaction leads to a Hubbard 
Hamiltonian 
\begin{equation}
H_{\rm HM} = - \sum_{ij \sigma} t_{ij} c_{i \sigma}^{\dag} c_{j 
\sigma} + U \sum_i c_{i \uparrow}^{\dag} c_{i \uparrow}
c_{i \downarrow}^{\dag} c_{i \downarrow} .
\label{ehh}
\end{equation}
Examining the stability of an antiferromagnetic state with 
wavevector ${\bf k}_M$, we note first that there will be 
no increase of the unit cell in the ($x,y$) plane, since there are already 
two atoms for each spin direction, but that it doubles along 
${\hat {\bf z}}$ when the spins order along the ladder legs.
In the notation of $A$ and $B$ sublattices for the bipartite 
system, sites $m = 1,3$ of the original unit cell (Fig. 1(c)) in every second 
plane, and $m = 2,4$ in the alternating planes, will 
belong to the $A$ sublattice, while the remaining sites will belong to 
$B$. Introducing the parameter $\delta n_A = n_{\uparrow} - n_{\downarrow}$
as the difference between average site occupation by particles of each spin 
orientation on a site of the $A$ sublattice, we require $\delta n_A = - 
\delta n_B = \delta n$ for all sites. We proceed by solving the problem in the
Hartree-Fock approximation, with the value of $U$ where $\delta n$ becomes 
finite marking the antiferromagnetic transition.

	Following the treatment of section II, the Hamiltonian may be written 
as in Eq. (\ref{esmh}), with now ${\bf c}_{{\bf k} \sigma}^{\dag} = 
\left(c_{{\bf k} \sigma}^{1a \dag}, c_{{\bf k} \sigma}^{1b \dag}, \dots 
\right)$, where the superscripts $a$ and $b$ denote atoms in the two ($x,y$) 
planes of the doubled unit cell, and 
\begin{equation}
{\bf H}_{\bf k} = - \left( \begin{array}{cccc}
{\bf M}_+ & {\bf S} (k_z) & {\bf 0} & {\bf {\overline R}}^* (k_z) \\
{\bf S}^* (- k_z) & {\bf M}_- & {\bf R} (k_z) & {\bf 0} \\
{\bf 0} & {\bf R}^* (- k_z) & {\bf M}_+ & {\bf S} (k_z) \\ 
{\bf {\overline R}} (- k_z) & {\bf 0} & {\bf S}^* (- k_z) & {\bf M}_- 
\end{array} \right) ,
\label{eshhm}
\end{equation}
in which the $2 \times 2$ matrices 
\begin{equation}
{\bf M}_{\pm} = \left( \begin{array}{cc}
2 t_{z}^{\prime} \cos k_z \pm \mbox{$\frac{1}{2}$} U \delta n & t_{z}
{\rm e}^{i \mbox{$\frac{1}{2}$} k_z} \\ t_{z} {\rm e}^{-i 
\mbox{$\frac{1}{2}$} k_z} & 2 t_{z}^{\prime} \cos k_z \mp 
\mbox{$\frac{1}{2}$} U \delta n \end{array} \right) ,
\label{ehmsm}
\end{equation}
\begin{equation}
{\bf R} (k_z) = \left( \begin{array}{cc}
t_r {\rm e}^{i {\bf k}.{\bf r}} & t_{r}^{\prime} {\rm e}^{i {\bf k}.{\bf r} 
+ i \mbox{$\frac{1}{2}$} k_z} \\ t_{r}^{\prime} {\rm e}^{i {\bf k}.{\bf r}
- i \mbox{$\frac{1}{2}$} k_z} & t_r {\rm e}^{i {\bf k}.{\bf r}} \end{array} 
\right) , 
\label{ehmsa}
\end{equation}
${\bf {\overline R}} (k_z)$ (defined identically using ${\bf {\overline r}}$)
and 
\begin{equation}
{\bf S} (k_z) = \left( \begin{array}{cc}
t_s \cos \mbox{$\frac{1}{2}$} k_x {\rm e}^{i k_y s_y} & t_{s}^{\prime} 
\cos \mbox{$\frac{1}{2}$} k_x {\rm e}^{i k_y s_y + i \mbox{$\frac{1}{2}$} k_z}
\\ t_{s}^{\prime} \cos \mbox{$\frac{1}{2}$} k_x {\rm e}^{i k_y s_y - i 
\mbox{$\frac{1}{2}$} k_z} & t_s \cos \mbox{$\frac{1}{2}$} k_x {\rm e}^{i 
k_y s_y} \end{array} \right) 
\label{ehmsb}
\end{equation}
are the generalizations of the previous expressions to the new unit cell, and 
the new
$c$-axis dimension is set to unity. This matrix cannot be block-diagonalized, 
but the structure of the solution for the eigenmodes and eigenvectors is 
evident from section II. Schematically, if the ($k_x, k_y$) dispersion
contained in the square root in Eq. (\ref{eqpb}) is denoted by ${\overline 
\epsilon}_{{\bf k}}^2$, the eight band dispersions will have the form 
\begin{equation}
E_{\bf k}^i = \pm \left[ {\overline \epsilon}_{{\bf k}}^2 + 
4 t_{z}^2 \cos^2 \mbox{$\frac{1}{2}$} k_z \pm \mbox{$\frac{1}{4}$} U^2 
\delta n^2 \right]^{1/2} - 2 t_{z}^{\prime} \cos 2 k_z ,
\label{eshmb}
\end{equation}
where $i$ labels the bands. The $U \delta n$ term splits the former four bands
into two sets, which as $U$ becomes large will not overlap, ensuring that the
half-filled system becomes insulating. The corresponding eigenvectors are
the states 
\begin{equation}
\psi_{{\bf k}}^i = \sum_{j=1}^{8} C_{j}^{i} \phi_{{\bf k}}^{j} ,
\label{eev}
\end{equation}
where the index $j$ runs over the eight atomic sites, and
$\phi_{{\bf k}}^{j}$ is the Bloch state created by the operator 
$c_{{\bf k}}^{j \dag}$. The Hartree-Fock equations for the system, which 
will determine
$\delta n$ and the chemical potential $\mu$ at fixed $U$ when solved 
self-consistently, are the equations for the total and site occupancies 
\begin{equation}
1 - \delta = \frac{1}{4} \sum_{i = 1}^{8} \int \frac{d^3 k}{(2 \pi)^3} 
\frac{1}{{\rm e}^{\beta \xi_{\bf k}^i} + 1}
\label{ehf1}
\end{equation}
and 
\begin{equation}
\mbox{$\frac{1}{2}$} \left( 1 - \delta \right) \left( 1 + \delta n \right) 
= \sum_{i = 1}^{8} \int \frac{d^3 k}{(2 \pi)^3} \frac{|C_{j}^{i}|^2}
{{\rm e}^{\beta \xi_{\bf k}^i} + 1} . 
\label{ehf2}
\end{equation}
These equations have been generalized to arbitrary band filling, which is
expressed in terms of the deviation $\delta$ from half-filling, which in turn
is normalized to be $1$. The chemical potential is contained in $\xi_{\bf k}^i 
= E_{\bf k}^i - \mu$, and the second expression is 
valid for the coefficients of any chosen Bloch function $j$. 

	The operation of diagonalizing the Hamiltonian matrix (\ref{eshhm})
can be performed numerically at sufficient speed that it is still possible to
solve the Hartree-Fock equations, which involve three-dimensional 
${\bf k}$-integration, on a workstation within a reasonable amount of time. 
In Fig. 5(a) are shown the energy bands for the incipient antiferromagnetic
system for half-filling ($\delta = 0$) and $U = t$. The splitting of the
bands into upper and lower branches as a result of $U$ is clearly evident, as
is the characteristic $\Gamma$XSY$\Gamma$ and ZURTZ structure of the
purely kinetic Hamiltonian in both sets of bands, at positive and negative
energies, as a result of the folding back of bands due to the change in 
meaning of the coordinate $k_z$, which now spans a Brillouin zone half the 
former size. At this value of $U$, the half-filled system would appear to be 
close to the transition from metallic to insulating behavior, which one 
expects near the point where there is no longer
any overlap between upper and lower band energies in any region of reciprocal
space. At smaller values of $U$, the two sets of bands are characteristic of 
the doubled Brillouin zone, and have a semi-metallic overlap, while as $U$ is
raised to large values, the energy gap increases, and the bands become
progressively more flat. 

	In other cuprate compounds, a Hubbard model has been found 
\cite{rhssj} to give an accurate description of the low-energy behavior 
with a ratio $U/W$ of order unity, where $W$ is the bandwidth. Here the total 
bandwidth of the uppermost valence bands is approximately the same as in 
the cuprates with CuO$_2$ planes, $W \simeq 3$eV. Since the local environment 
of the Cu$^{2+}$ ions is similar, the on-site Coulomb repulsion should also 
be the same, $U \simeq 4$eV. These parameter values place LaCuO$_{2.5}$ well 
within the Mott insulating region. We note that the actual magnetic structure 
cannot be determined in the Hartree-Fock Approximation, as the quantum 
corrections which act to stabilize the spin liquid phase are not included. 

	The form of the Hartree-Fock solutions for smaller values of $U$ are 
sensitive to the effective dimensionality of the electronic structure. In 
general, values of $U \sim W$ are required to obtain a Mott insulator at 
half-filling, but the perfect-nesting property of a one-dimensional band 
gives an insulating state for arbitrarily small values of $U$. In Fig. 5(b), 
the average antiferromagnetic order (parameterized by $\delta n$) in the 
half-filled system is shown as a function of the on-site repulsion $U$, 
which is measured in terms of the bandwidth $W \equiv 3$eV. In 
LaCuO$_{2.5}$ we find that the critical value $U_c$ for the antiferromagnetic 
transition is rather small, $U_c \simeq 0.2 W$, which indicates substantial 
one-dimensional nature. The interladder hopping matrix elements, although 
quite strong, are not sufficient to destroy the nesting character completely, 
which would make the electronic structure effectively three-dimensional. 

	This quasi-one dimensional behavior of the bands may also be reflected 
in the sensitivity of the system to random potential fluctuations. It is well 
known that the onset of localization is strongly dependent on dimensionality. 
Clearly, in ladder compounds there is an inherent conflict between the need 
to change the valence of the counterions in order to induce hole carriers, 
and the need to avoid strong, random potential fluctuations. Thus it appears 
that in La$_{1-x}$Sr$_x$CuO$_{2.5}$ the random potential fluctuations act to 
cause localization for $x \le 0.15$, in spite of the substantial interladder 
overlap, suggesting that a more gentle hole-doping technique will be 
required to retain itinerant character at small doping concentrations. This 
could perhaps be achieved in structures where the counterions are further 
from the ladders than in the present case of LaCuO$_{2.5}$.

\section{Conclusion}

	We have investigated the basic electronic and magnetic properties of
the three-dimensionally coupled 2-leg ladder compound LaCuO$_{2.5}$, a
material which is of significant experimental and theoretical interest as it
is the first ladder compound to be discovered in which the Cu$_2$O$_3$ ladders
may be doped with holes. We present a tight-binding fit in which the bands are
derived from a single ($d_{x^2 - y^2}$) orbital close to the Fermi energy on 
each copper atom. The results of LDA studies are well reproduced by reasonable
values of the most significant transfer integrals: those within each ladder
are found to be similar to the CuO$_2$ planar system, emphasizing the 
short-range nature of the dominant physical processes, while the 
interladder hopping term $t^{\prime} \simeq \; $\mbox{$\frac{1}{2}$}$
t$ is found to be quite large. As a consequence, the interladder
magnetic coupling $J^{\prime}$ is also relatively large, and the compound may
be expected to exhibit some three-dimensional characteristics. 

	The effective spin interactions in this structure are those of 
unfrustrated antiferromagnetism, and a mean-field treatment of the magnetic 
state from the basis of dimerized singlets on the rungs of decoupled ladders 
gives a spin-liquid phase whose spin gap decreases with increasing interladder 
coupling. The spin gap is found to vanish, signalling a transition to an 
ordered phase, at an interladder coupling ratio $J^{\prime}/J$ somewhat
smaller than that deduced for LaCuO$_{2.5}$, indicating that the system is 
located in the antiferromagnetically ordered state, albeit not far from 
the quantum critical point of the ordering transition. We may take the very 
small value of the intrinsic susceptibility measured at low temperatures
\cite{rht} as evidence that proximity to the critical point, and the
possibility this allows of significant critical fluctuations, plays an
important role in determining the physics of the system. 

	Further
insight into the electronic properties is provided by the introduction of an 
on-site repulsion term, $U$, to the band scheme: within the Hartree-Fock 
approximation we find that in the half-filled system the transition to an 
antiferromagnetic insulator occurs for values of $U$ quite small compared to 
the bandwidth. This is a reflection of the fact that the predominant feature
of the bands remains the dispersion in the ladder direction, and this
one-dimensionality, also apparent in the Fermi surfaces of the
partially-filled bands, makes the system inherently susceptible to potential 
fluctuations. That $U_c$ is finite illustrates an absence of perfect nesting 
conditions in the bandstructure, and the instability occurs at the 
antiferromagnetic wavevector.

	Our results represent the first theoretical consideration of the
LaCuO$_{2.5}$ system. While by the nature of the approximate techniques used
they are somewhat inexact, we believe that they are important in establishing
the parameter space for the magnetic state, and in focussing the direction of
further research. The properties of such ``nearly critical'' magnetic systems 
have been studied for the two-dimensional, planar case by detailed analytical
\cite{rcsy} and numerical \cite{rtpc} techniques. 
We propose the application of similar methods for the case of ladder systems 
with variable interladder coupling in two additional dimensions, to 
investigate the nature of the critical point, and the surrounding ``quantum 
critical'' regime, in the same formalism, with particular emphasis on the 
appearance of physical quantities accessible in experiment. Such studies may
prove valuable in establishing the framework within which to interpret the
data from experiment, and hopefully will serve to reconcile the apparent
contradiction in current results.

	The observation that the spin liquid is not the appropriate
description of the magnetic ground state in LaCuO$_{2.5}$ is itself 
important, particularly with regard to the current interpretation of the 
susceptibility 
data.\cite{rht} While it may be possible to deduce the form of the 
susceptibility suitable for the critical regime from a mean-field picture, 
we await the
results of numerical studies of the same system in order to make a more
detailed comparison with the data. A further direction not strongly emphasized
in this communication is the nature of the doped system: the methods used
here become less accurate at finite doping, and so were not studied in great
detail in this regime, but may nonetheless be used to obtain additional 
insight into the evolution of spin properties on moving towards the metallic 
state.

	We are grateful to Y. Kitaoka, M. Sigrist, M. Takano, M. Troyer and 
especially Z. Hiroi and M. Zhitomirsky for helpful discussions, and to the 
Swiss National Fund for financial support.

\eject
\section*{Figures}

\begin{figure}
\caption{(a) Crystal stucture of the depleted perovskite LaCuO$_{2.5}$. Black
and white spheres within the ladder units represent Cu and O atoms, 
respectively, and grey spheres represent Sr. The material is viewed along 
the axis of the ladders (${\hat {\bf z}}$), which can be seen to
be rotated about this axis to a relative angle of 62$^o$. In any ($x,y$) plane
each copper atom is bonded to a neighbor in the same ladder by a rung bond,
and to two copper atoms in neighboring ladders. The interladder couplings are
through one oxygen atom which is bonded as part of the square planar
coordination in the same ladder and is apical to a copper atom in the next 
ladder, and through the single apical oxygen atom, which is bonded in the 
ladder of the other neighbor. Allocation of antiferromagnetically arranged
spins to each ladder shows that the material can exist as an unfrustrated
antiferromagnet. (b) Schematic representation of LaCuO$_{2.5}$ to show 
the tight-binding parameters between Cu atoms used to fit the LDA 
bandstructure. (c) Appearance of the four inequivalent Cu atoms (black 
circles) in the unit cell. White circles represent O atoms. The vectors for 
the two types of bond in the 
($x,y$) plane are ${\bf r} ({\bf {\overline r}}) = (0.5856a,- (+) 0.2114b)$ 
and ${\bf s} ({\bf {\overline s}}) = (+ (-) 0.5a, 0.2886b)$, where 
$a \simeq \protect{\sqrt{2}} a_p$ and
$b \simeq 2 \protect{\sqrt{2}} a_p$ are the lattice constants in the 
$x$ and $y$ directions, and $a_p = c$ is the lattice constant 
of the original, cubic perovskite structure.\protect{\cite{rht}}}

\end{figure}

\begin{figure}
\caption{
(a) Illustration of the effects of the chosen bandstructure parameters
on the observed dispersion curves, for the choices $t_z = t_r = t = 
0.4$eV, $t_{z}^{\prime} = \, $\mbox{$\frac{1}{6}$}$ t_z$, $t_{r}^{\prime} = - 
$\mbox{$\frac{1}{5}$}$ t_r$, $t_s = \, $\mbox{$\frac{2}{5}$}$t_r$ and 
$t_{s}^{\prime} = \, $\mbox{$\frac{1}{5}$}$ t_s$. (b) Tight-binding 
bandstructure for the parameter set which appears closest to the LDA results 
of Mattheiss.\protect{\cite{rm}} Parameters are as in (a), but with $t_s = \, 
$\mbox{$\frac{1}{2}$}$ t_r$. Details of the fitting procedure and parameter 
choices are given in the text. (c) Notation for ${\bf k}$ points in the 
orthorhombic Brillouin zone.}
\end{figure}

\begin{figure}
\caption{
Schematic representation of the periodic structure in the ($x,y$) 
plane of locally antiferromagnetically-correlated spins, for calculation of 
the spinon dispersion. The magnetic interaction parameters shown are $J$ on 
the ladder rungs and $J^{\prime} \equiv \lambda^{\prime} J$ between spins 
on neighboring ladders.
}
\end{figure}

\begin{figure}
\caption{
(a) Spin gap $\Delta$ as a function of the ratio between interladder 
and intraladder magnetic couplings $J^{\prime} / J$, calculated for a
three-dimensional system at $T = 0$, in 
the mean-field approximation from the starting point of dimerized ladder 
rungs. Here the intraladder rung and leg interactions are the same ($\lambda 
= 1$), and the value of the spin gap at $J^{\prime} = 0$ is that of the
noninteracting, 2-leg ladder ($\Delta_0 = 0.501 J$) in the same approximation.
(b) Spin gap for a two-dimensional array of ladders.  }
\end{figure}

\begin{figure}
\caption{
(a) Bandstructure computed in the Hartree-Fock approximation for 
the parameters of the tight-binding fit of section II, with the inclusion of 
an on-site repulsion parameter $U$. The bands are shown for $U = t = 0.4$eV, 
where the half-filled system is close the metal-insulator transition. 
(b) Antiferromagnetic order as a function of $U$. The degree of ordering is 
parameterized by $\delta n = n_{\uparrow} - n_{\downarrow}$, the
difference between the average occupation of each site in the structure
by particles with spins oriented upwards and downwards. The doping is 
$\delta = 0$, i.e. the system is half-filled, and the temperature is taken to
be low. 
}
\end{figure}


\begin{thebibliography}{99}
\bibitem{mzhc} We are indebted to M. E. Zhitomirsky for pointing out to 
us an error in the original version of the following calculation.
\end{thebibliography}
\end{document}